\documentclass[a4paper]{article}
\def\be{\begin{equation}}
\def\eea{\end{eqnarray}}
\def\bea{\begin{eqnarray}}
\def\ee{\end{equation}}
\def\l{\lambda}

\def\d{\mathrm{d}}
\def\a{\alpha}

\usepackage[psamsfonts]{amssymb}
\usepackage{amsmath}
\usepackage{cite}

\author{M. Alimohammadi\footnote{alimohmd@ut.ac.ir}\ \ and H.
Mohseni Sadjadi
\\ {\small School of Physics, University of Tehran,}
\\ {\small North Karegar Ave., Tehran, Iran.}}
\title{ Attractor solutions for general hessence dark energy}
\begin{document}
\maketitle
\begin{abstract}

\noindent As a candidate for the dark energy, the hessence model
has been recently introduced. We discuss the critical points of
this model in almost general case, that is for arbitrary hessence
potential and almost arbitrary hessence-background matter
interaction. It is shown that in all models, there always exist
some stable late-time attractors. It is shown that our general
results coincide with those solutions obtained earlier for special
cases, but some of them are new. These new solutions have two
unique characteristics. First the hessence field has finite value
in these solutions and second, their stabilities depend on the
second derivative of the hessence potential.
\end{abstract}

\section{Introduction}
In recent years, astronomical observations from type Ia supernova
\cite{1}, WMAP data \cite{2}, and large scale structure surveys
\cite{3}, have shown that the expansion of the universe is
accelerated. Although there is no clear understanding of the
mechanism leading to this acceleration, but it is believed that
about $70\%$ of the total energy density of universe consists of
this unknown energy, i.e. dark energy, which leads to this
expansion. The simplest explanation of dark energy is a
cosmological constant $\Lambda$ of order ($10^{-3}$ ev)$^4$.
Unfortunately it is about 120 orders smaller than the naive
expectations, gives rise to the idea of a dynamical nature of this
energy. The possible dynamical explanations have been introduced
in different frameworks, such as quintessence \cite{4}, phantom
\cite{5}, k-essence \cite{6}, tachyons \cite{7}, etc.

In studying the dark energy, the equation of state parameter
$w_{\rm de}=p_{\rm de}/\rho_{\rm de}$ plays an important role,
where $p_{\rm de}$ and $\rho_{\rm de}$ are the pressure and energy
density of the dark energy, respectively. This parameter is always
equal to constant -1 in cosmological constant model, but it can be
a dynamical variable in the above mentioned dynamical models. This
is an important point since the present data seems to slightly
favor an evolving dark energy with $w_{\rm de}$ being below -1
around present epoch, \cite{8}, from $w_{\rm de}>-1$ in the near
past \cite{9}.

To be definite, we consider the following action
 \be\label{1}
 S=\int \d^{4}x\sqrt{-g}\left(-\frac{\cal R}{16\pi G}+{\cal L}_{DE}+{\cal
L}_m\right),
 \ee
where $g$ is the determinant of the metric $g_{\mu\nu}$, $\cal R$
is the Ricci scalar, ${\cal L}_{DE}$ and ${\cal L}_m$ are the
Lagrangian densities of the dark energy and matter, respectively.
In the case of quintessence, the Lagrangian density is
\be\label{2}
 {\cal
L}_{\rm
quintessence}=\frac{1}{2}\left(\partial_{\mu}\phi\right)^2-V(\phi),
 \ee
 where $\phi$ is a real scalar field. In a
spatially flat Friedmann-Robertson-Walker (FRW) universe with
homogeneous $\phi$, $w$ is
 \be\label{3}
w_{\rm quintessence}=\frac{\dot{\phi}^{2}/2-V(\phi)}
{\dot{\phi}^{2}/2 +V(\phi)}.
 \ee
which results $-1\leq w_{\rm quintessence}$. The Lagrangian
density of phantom scalar field is
 \be\label{4}
 {\cal
L}_{\rm
phantom}=-\frac{1}{2}\left(\partial_{\mu}\phi\right)^2-V(\phi),
 \ee
from which
 \be\label{5}
w_{\rm phantom}=\frac{-\dot{\phi}^{2}/2-V(\phi)}
{-\dot{\phi}^{2}/2 +V(\phi)},
 \ee
where for $\rho \geq 0$, which comes from $H^2=(8\pi G/3)\rho$, it
results $w_{\rm phantom} \leq -1$. So we can not cross the
phantom-divide-line $w=-1$ in quintessence or phantom model alone.
A possible way to overcome this problem is considering two real
fields, which one behaves as quintessence and other one as phantom
field. The resulting model, called the quintom model, has the
following Lagrangian \cite{9,10}
 \be\label{6}
 {\cal
L}_{\rm quintom}=\frac{1}{2}\left(\partial_{\mu}\phi_1\right)^2-
\frac{1}{2}\left(\partial_{\mu}\phi_2\right)^2-V(\phi_1,\phi_2),
 \ee
with
 \be\label{7}
w_{\rm
quintom}=\frac{\dot{\phi_1}^{2}/2-\dot{\phi_2}^{2}/2-V(\phi_1,\phi_2)}
{\dot{\phi_1}^{2}/2-\dot{\phi_2}^{2}/2 +V(\phi_1,\phi_2)}.
 \ee
Now it is obvious that $w_{\rm quintom}\geq -1$ when
$\dot{\phi_1}^{2}\geq \dot{\phi_2}^{2}$ and  $w_{\rm quintom}< -1$
when $\dot{\phi_1}^{2}< \dot{\phi_2}^{2}$. So crossing the
phantom-divide-line is, in principle, possible in quintom model.
See, for example, \cite{10} and \cite{11}.

Instead of introducing two independent real scalar field to
describe a quintom model, it is also natural to consider a single
complex scalar field. The resulting spintessence model of dark
energy \cite{12,13,14,15,n1}, has the following Lagrangian density
  \be\label{8}
 {\cal L}_{\rm
spintessence}=\frac{1}{2}\left(\partial^{\mu}\Phi^*\right)
\left(\partial_{\mu}\Phi\right)-V(|\Phi |).
 \ee
Using $\Phi =\phi_1+i\phi_2$, the kinetic term of eq.(\ref{8})
reduces to the kinetic terms of eq.(\ref{6}). Also the above
Lagrangian is invariant under $\Phi \rightarrow e^{i\alpha}\Phi$
which leads to a conserved charge. Unfortunately this model
suffers from the problem of Q-ball formation \cite{12,14}. Q-ball
is a kind of nontopological soliton which except in some special
cases of spintessence with unnatural potentials, grows
exponentially and depends on the potential, can be either stable
at the late-time to be a dark matter, or decay into other
particles. Therefore the spintessence model can not be a viable
candidate for the dark energy.

To avoid the difficulty of Q-ball formation and also to introduce
another possibility for mysterious dark energy problem, a
non-canonical complex scalar field, called hessence, has been
recently introduced in \cite{16}. In the hessence model, the
phantom-like role is played by the so called internal motion
$\dot\theta$, where $\theta$ is the internal degree of freedom of
hessence. There is a conserved charge $Q$ in this model which
makes the physics of hessence more interesting, and the transition
from $w_h>-1$ to $w_h<-1$ or vice versa is also possible. Another
interesting feature of hessence model is that it is free of
big-rip\cite{17}. If $w<-1$ in an expanding FRW universe, then the
positive energy density of a phantom matter generally becomes
infinite in finite time, overcoming all other forms of matter and
hence leads to the late-time singularity called the ''big-rip''
\cite{18}.

By considering two specific hessence potentials, i.e. the
exponential and the (inverse) power law, and four different
interaction forms between hessence and background perfect fluid,
the late-time attractors of hessence model have been studied in
\cite{17}. In each case, different scaling and hessence-dominated
solutions have been obtained and their stability properties have
been studied.

In this paper we are going to study the late-time attractors of
the almost general hessence model, with arbitrary hessence
potential and almost arbitrary hessence-background matter
interaction term. By almost arbitrary, we mean that the hessence
potential and hessence-background matter interaction terms are
arbitrary functions of dimensionless variables defined in
(\ref{22}). Specifically, we mean eqs.(\ref{n1}) and (\ref{n2}).
We show that there always exist some stable solutions, in scaling
or hessence-dominated form, which some of them have not been
appeared in special cases studied in \cite{17}. These new
solutions have two interesting unique characteristics which are
absent in other solutions. First their stability depends on the
second derivative of the hessence potential, and second the
hessence field has finite value in these solutions. The
significance of the second derivative of the potential in the
late-time behaviors has been also revealed for quintessence model
in \cite{rendall}, in which some conditions have been imposed on
the first and second derivatives of the potential.

The scheme of the paper is as follows. In section 2, we briefly
introduce the main points of hessence model and the system of
equations which determines the critical points in terms of
dimensionless variables. In section 3, we consider the general
hessence potential, but assuming no hessence-background matter
interaction. It is shown that there exists five general solutions
for critical points which three of them are stable under specific
conditions. Finally in section 4, the hessence potential and
hessence-background matter interaction are considered arbitrary
and it is shown that there are, in general, six classes of
solutions for critical points. The stability of solutions is
discussed in special cases. It is shown that all the solutions of
\cite{17} can be obtained from our general results.

We use the units $\hbar=c=1,$ $\kappa^2=8\pi G$ and adopt the
metric convention as $(+,-,-,-)$ throughout the paper.

\section{Hessence model}
Following \cite{16}, the hessence field introduced by a
non-canonical complex scalar field
 \be\label{9}
 \Phi =\phi_1+i\phi_2,
 \ee
with Lagrangian density
 \be\label{10}
  {\cal L}_h=\frac{1}{4}\left[\,(\partial_\mu \Phi)^2+ (\partial_\mu
\Phi^\ast)^2\,\right]-U(\Phi^2+ \Phi^{\ast 2})=\frac{1}{2}\left[\,
(\partial_\mu \phi)^2-\phi^2
(\partial_\mu\theta)^2\,\right]-V(\phi),
 \ee
 where the new fields $(\phi,\theta)$ are defined through
 \be\label{11}
\phi_1=\phi\cosh\theta,~~~~~~~\phi_2=\phi\sinh\theta.
 \ee
In a spatially flat FRW universe with scale factor $a(t)$, the
equations of motion for $\phi$ and $\theta$, when they are
considered homogeneous, are
 \bea
\ddot{\phi}+3H\dot{\phi}+\phi\dot{\theta}^2+V_{,\phi}=0, \label{12}\\
\phi^2\ddot{\theta}+(2\phi\dot{\phi}+3H\phi^2)\dot{\theta}=0,\label{13}
 \eea
where $H\equiv\dot{a}/a$ is the Hubble parameter and overdot and
subscript ``$_{,\phi}$'' denote the derivatives with respect to
cosmic time $t$ and $\phi$, respectively.  Eq.(\ref{13}) implies
 \be\label{14}
  Q=a^3 \phi^2\dot{\theta}={\rm const.},
  \ee
where $Q$ is the total conserved charge due to the symmetry of
Lagrangian (\ref{10}) under the transformation
$\phi\rightarrow\phi$ and $\theta\rightarrow\theta -i\alpha$.
Substituting eq.(\ref{14}) into (\ref{12}), one has
 \be\label{15}
\ddot{\phi}+3H\dot{\phi}+\frac{Q^2}{a^6\phi^3}+V_{,\phi}=0.
 \ee
The pressure and energy density of hessence are
 \be\label{16}
p_h=\frac{1}{2}\dot{\phi}^2-\frac{Q^2}{2a^6\phi^2}-V(\phi),~~~~~~~
\rho_h=\frac{1}{2}\dot{\phi}^2-\frac{Q^2}{2a^6 \phi^2}+V(\phi).
 \ee
The Friedmann equation and Raychaudhuri equation are given by,
respectively,
 \bea
&&H^2=\frac{\kappa^2}{3}\left(\rho_h+\rho_m\right),\label{17}\\
&&\dot{H}=-\frac{\kappa^2}{2}\left(\rho_h+\rho_m+p_h+p_m\right),\label{18}
 \eea
where $p_m$ and $\rho_m$ are the pressure and energy density of
background matter, respectively. The background matter is
described by a perfect fluid with barotropic equation of state
 \be\label{19}
 p_m=w_m\rho_m\equiv (\gamma-1)\rho_m,
 \ee
where $0<\gamma<2$. In particular, $\gamma=1$ and $4/3$ correspond
to dust matter and radiation, respectively.

To introduce the interaction between hessence and background
matter, it is assumed that it can be described by an interaction
term $C$ in the energy balance \cite{17,19}
  \bea
\dot{\rho}_h+3H\left(\rho_h+p_h\right)=-C,\label{20}\\
\dot{\rho}_m+3H\left(\rho_m+p_m\right)=C,\label{21}
 \eea
which preserves the total energy conservation equation
$\dot{\rho}_{tot}+3H\left(\rho_{tot}+p_{tot}\right)=0$. $C=0$
corresponds to no interaction between hessence and background
matter and when $C\not=0$, a new term due to $C$ will appear in
the right hand side of eq.(\ref{15}).

Following \cite{20} and many other papers, if we introduce the
following dimensionless variables
 \be\label{22}
x\equiv\frac{\kappa\dot{\phi}}{\sqrt{6}H},~~~y\equiv\frac{\kappa\sqrt{V}}{\sqrt{3}H},~~~
z\equiv\frac{\kappa\sqrt{\rho_m}}{\sqrt{3}H},~~~u\equiv\frac{\sqrt{6}}{\kappa\phi},~~~
v\equiv\frac{\kappa}{\sqrt{6}H}\frac{Q}{a^3 \phi},
 \ee
then using eqs.(\ref{16})-(\ref{18}),(\ref{20}), and (\ref{21}),
the evolution equations of these variables become
 \bea
&& x^\prime=3x\left(x^2-v^2+\frac{\gamma}{2}z^2-1\right)-uv^2
-\sqrt{\frac{3}{2}}y^2f-C_1,\label{23}\\
&& y^\prime=3y\left(x^2-v^2+\frac{\gamma}{2}z^2\right)
+\sqrt{\frac{3}{2}}x y f,\label{24}\\
&& z^\prime=3z\left(x^2-v^2+\frac{\gamma}{2}z^2-\frac{\gamma}{2}\right)+C_2,\label{25}\\
&& u^\prime=-xu^2,\label{26}\\
&&
v^\prime=3v\left(x^2-v^2+\frac{\gamma}{2}z^2-1\right)-xuv.\label{27}
 \eea
Prime denotes derivative with respect to the e-folding time ${\cal
N}\equiv\ln a$, and
 \bea
 &&f\equiv\frac{V_{,\phi}}{\kappa V},\label{28}\\
  C_1\equiv\frac{\kappa C}{\sqrt{6}H^2
\dot{\phi}},&& ~~~~C_2\equiv\frac{\kappa
C}{2\sqrt{3}H^2\sqrt{\rho_m}}=\frac{x}{z}C_1. \label{29}
 \eea
The Friedmann equation (\ref{17}) becomes
 \be\label{30}
 x^2+y^2+z^2-v^2=1,
 \ee
and the fractional energy densities are
  \be\label{31}
  \Omega_h=\frac{\rho_h}{\rho_c}=x^2+y^2-v^2,~~~~~~~ \Omega_m=\frac{\rho_m}{\rho_c}=z^2,
 \ee
where $\rho_c=\frac{3H^2}{\kappa}$ is the critical energy density.
The equation of states of hessence and whole system are
 \be\label{32}
w_h=\frac{p_h}{\rho_h}=\frac{x^2-v^2-y^2}{x^2-v^2+y^2},~~~~~~~
w_{eff}=\frac{p_h+p_m}{\rho_h+\rho_m}=x^2-v^2-y^2+(\gamma-1)z^2.
 \ee
The critical points $(\bar{x},\bar{y},\bar{z},\bar{u},\bar{v})$
are obtained by imposing the conditions
$\bar{x}^\prime=\bar{y}^\prime= \bar{z}^\prime
=\bar{u}^\prime=\bar{v}^\prime=0$.

Note that we take
 \be\label{n1}
 f=f(u),
 \ee
and
 \be\label{n2}
 C_1=C_1(x,z,u),~~~~~C_2=C_2(x,z,u).
 \ee
It is because $V=V(\phi)$, so $f$ is assumed to be a function of
only one variable $u$. Also because of eq.(\ref{30}), only four of
the variables (\ref{22}) are independent, which we can take them
$x,$ $y$, $z$, and $u$. But the dependence of hessence-background
matter interaction to the potential $V$ (or variable $y$) is
meaningless, so $C$s are taken to be arbitrary functions of
variables $x$, $z$ and $u$.  Really eq.(\ref{n1}) does not
constraint the potential $V$, but eq.(\ref{n2}) restricts the
possible interaction term $C$. In this way the
eqs.(\ref{23})-(\ref{27}) become autonomous and we need not
consider any further variables. For example $f$ is not and extra
variable since $f'=0$ leads to, for an arbitrary potential,
$u'=0$.

\section{Attractors in $C=0$ case}
To obtain the attractors for arbitrary hessence potential and when
there is no hessence-background matter interaction, we must solve
eq.(\ref{30}) and the set of equations (\ref{23})-(\ref{27}), when
setting zero, in $C_1=C_2=0$. Eq.(\ref{26}) results $\bar{u}=0$ or
$\bar{x}=0$ and eq.(\ref{25}) results $\bar{z}=0$ or
$\bar{x}^2-\bar{v}^2+(\gamma /2)\bar{z}^2-\gamma /2=0$. So we have
four possibilities: I$=(\bar{u}=0,\bar{z}=0)$, II$=(\bar{u}=0,
\bar{x}^2-\bar{v}^2+(\gamma /2)\bar{z}^2-\gamma /2=0)$, III$=
(\bar{x}=0,\bar{z}=0)$ and IV$=(\bar{x}=0,
\bar{x}^2-\bar{v}^2+(\gamma /2)\bar{z}^2-\gamma /2=0)$. In type I
solution, eq.(\ref{27}), using (\ref{30}), reduces to $vy^2=0$, so
it divides to I.1$=(\bar{u}=0,\bar{z}=0,\bar{v}=0)$ and
I.2$=(\bar{u}=0,\bar{z}=0,\bar{y}=0)$ solutions. The remaining
variables can be easily found. The final results are represented
in Table \ref{tab1}.
\begin{table}[htbp]
\begin{center}
\begin{tabular}{c|c|cccc}
\hline\hline Label & Critical Point
$(\bar{x},\bar{y},\bar{z},\bar{u},\bar{v})$ & $\Omega_h$ &
$\Omega_m$ & $w_h$ & $w_{eff}$\\ \hline
 I.1 & $-\frac{\bar{f}}{\sqrt{6}}$, \
 $\sqrt{1-\frac{{\bar{f}}^2}{6}}$,\ 0,\ 0,\ 0& 1 & 0 &$-1+\frac{{\bar{f}}^2}{3}$&$-1+\frac{{\bar{f}}^2}{3}$\\
I.2 & $\bar{x}^2\geq 1$,\ 0,\ 0,\ 0,\ $\pm\sqrt{\bar{x}^2-1}$  & 1 & 0 & 1 & 1\\
II&$-\sqrt{\frac{3}{2}}\frac{\gamma}{\bar{f}}$,$\sqrt{\frac{3\gamma}{{\bar{f}}^2}(1-\frac{\gamma}{2})}$,
$\sqrt{1-\frac{3\gamma}{{\bar{f}}^2}}$,\ 0,\
0&$\frac{3\gamma}{{\bar{f}}^2}$&$1-\frac{3\gamma}{{\bar{f}}^2}$&$-1+\gamma$&$-1+\gamma$\\
III & 0,\ 1,\ 0,\ $\bar{f}=0$,\ 0 & 1 & 0 & -1 & -1\\
IV & 0,\ 0,\ 1,\ any,\ 0 & 0 & 1 & any & $-1+\gamma$\\
 \hline
\end{tabular}
\end{center}
\caption{\label{tab1} Critical points for arbitrary hessence
potential when there is no hessence-background matter
interaction.}
\end{table}

The solutions with $\Omega_h=1$, i.e. solutions I.1, I.2 and, III,
are hessence-dominated, solution IV is background-matter-dominated
and solution II is scaling solution. In solution III, $\bar{u}$
must be found by solving $\bar{f}\equiv f(\bar{u})=0$. Note that
in all cases, $w_{eff}>-1$. It is also interesting that for
potentials where $f^2(\bar{u}=0)=3\gamma$, the solutions I.1 and
II become degenerate.

In examples considered in \cite{17}, the potentials are
 \bea
 &&V_1=V_0e^{-\lambda \kappa\phi},\label{33}\\
 &&V_2=V_0(\kappa\phi)^n,\label{34}
 \eea
or, using (\ref{28}),
 \bea
 &&f_1=-\lambda,\label{35}\\
 &&f_2=\frac{n}{\kappa\phi}=\frac{nu}{\sqrt{6}}.\label{36}
 \eea
It can be easily checked that our solutions (I.1, I.2, II and IV)
and (I.1, I.2, and IV) reduce to those obtained in \cite{17} for
$V_1$ and $V_2$, respectively. The solution II does not exist for
potential $V_2$ since $f_2(\bar{u}=0)=0$. The solution III is a
new solution which has not been appeared in \cite{17}. This is
because the equation $\bar{f}=0$ results $\lambda =0$ for $V_1$,
which is not acceptable, and results $\bar{u}=0$ for $V_2$, which
reduces solution III to I.1.

To study the stability of the critical points I.1-IV, we must
consider a small perturbation about the critical point
$(\bar{x},\bar{y},\bar{z},\bar{u},\bar{v})$: $x\to\bar{x}+\delta
x$, $y\to\bar{y}+\delta y$, $z\to\bar{z}+\delta z$,
$u\to\bar{u}+\delta u$, and $v\to\bar{v}+\delta v$, in
eqs.(\ref{23})-(\ref{26}) with $C_1=C_2=0$, which due to Friedmann
constraint (\ref{30}), only four of them are independent. In this
way one can found a $4\times 4$ matrix $M$ defined through
 \be\label{37}
 \frac{\d}{\d {\cal N}}\left( \begin{array}{c} \delta q_1\\ \delta q_2\\
 \delta q_3\\ \delta q_4 \end{array} \right)=M\left( \begin{array}{c} \delta
  q_1\\ \delta q_2\\ \delta q_3\\ \delta q_4 \end{array} \right),
 \ee
where $(q_1,\cdots ,q_4)$ are four chosen independent variables.
The critical solutions are stable if the real part of all the
eigenvalues of matrix $M$ are negative. The eigenvalues of matrix
$M$ for our solutions are as following:
 \bea
 &&{\rm I.1}:(-3{\bar{y}}^2,-6{\bar{y}}^2,\frac{1}
 {2} ({\bar{f}}^2-3\gamma),0),\label{38} \\
 &&{\rm I.2}:(0,0,\frac{3}{2}(2-\gamma ),
 3+\sqrt{\frac{3}{2}}\bar{x}\bar{f}),\label{39} \\
 {\rm III}:( -6,-\frac{3\gamma}{2}&,&
 -\frac{3}{2}+ \frac{\sqrt{9+2\sqrt{6}{\bar{u}}^2\bar{f'}}}{2},
  -\frac{3}{2}- \frac{\sqrt{9+2\sqrt{6}{\bar{u}}^2\bar{f'}}}{2}
),\label{40} \\
 &&{\rm IV}:\left(0,\frac{3\gamma}{2},3(\gamma -2 ),
 \frac{3}{2}(\gamma -2 )\right),\label{41}
 \eea
in which $(q_1,\cdots ,q_4)=(x,y,z,u)$ and $\bar{f'}=\left( \d
f/\d u\right)_{\bar{u}}$. It is clear that I.1 solution is {\it
stable} if
 \be\label{42}
 {\bar{f}}^2\leq 3\gamma ,
 \ee
I.2 solution is {\it unstable} since $\gamma <2$, III solution is
{\it stable} if
 \be\label{43}
 \left( \frac{\d f}{\d u}\right)_{\bar{u}}\leq 0,
 \ee
and IV is an {\it unstable} solution  since $\gamma >0$.

For solution II, it is easier to use $(q_1,\cdots
,q_4)=(x,z,u,v)$. Then it can be easily found that $\l_1=0$,
$\l_2=\frac{3}{2}(\gamma -2 )$, and $\l_3$ and $\l_4$ are roots of
equation $\l^2+b\l+c=0$ with
 \bea\label{44}
 b&=&\frac{3}{2}(\gamma -2 ), \cr
 c&=&\frac{9\gamma}{2{\bar{f}}^2}(2-\gamma )({\bar{f}}^2-3\gamma
 ).
 \eea
$\l_3$ and $\l_4$ are non-positive if $b\geq 0$ and $c\geq 0$. As
$\gamma <2$, the solution II is {\it stable} if
 \be\label{45}
 {\bar{f}}^2\geq 3\gamma.
 \ee
So for any potential $V$, there {\it always} exists at least one
stable attractor. A hessence-dominated attractor (I.1) if
${\bar{f}}^2\leq 3\gamma$ or a scaling attractor (II) if
${\bar{f}}^2\geq 3\gamma$.

It is interesting to note that the solution III has two unique
properties. First, it is the only stable attractor which has the
non-vanishing $\bar{u}$ value, i.e. finite value of hessence field
$\bar{\phi}$. Second, it is the only attractor which its stability
depends on the derivative of $f$ (eq.(\ref{43})). In other words,
the stable attractors I.1 and II can not distinguish between
different potentials with the same $\bar{f}$ value, but the
attractor III does.

As examples of potentials which have the solution III as a stable
attractor, we may consider $V_3=V_0\sin(\kappa\phi/\sqrt{6})$ and
$V_4=V_0\cos(\kappa\phi/\sqrt{6})$. For $V_3$, we have
$f_3=-(1/\kappa u^2)\cot(1/u)$, which results $\bar{u}=[(2n+1)\pi
/2]^{-1}$, with $n=0,1,2,\cdots$, as the solution of $\bar{f_3}=0$
equation. It really has infinite number of attractors. Then $(\d
f/\d u)_{\bar{u}}=-(1/\kappa)[(2n+1)\pi /2]^4<0$, which shows that
the attractors are stable. The same is true for $V_4$ potential.

\section{Attractors in the presence of hessence-background matter
interaction}
 In this case, we must solve eq.(\ref{30}) and the set
of equations (\ref{23})-(\ref{27}), when setting zero, for
arbitrary $C$ function. Eq.(\ref{26}) results $\bar{u}=0$ or
$\bar{x}=0$. In each case, we consider eight cases in which each
of the variables $\bar{y}$, $\bar{z}$ and $\bar{v}$ has two
possibilities, zero and not zero, and then check the consistency
of the equations. The final results are as following:
 \be\label{46}
 {\rm solution \ \ 1}:\{ \bar{x}=-\frac{\bar{f}}{\sqrt{6}},\
 \bar{y}=\sqrt{1-\frac{{\bar{f}}^2}{6}},\ \bar{z}=\bar{u}=
 \bar{v}=0, \ \bar{C_1}=\bar{C_2}=0\}.
 \ee
Note that the equations $\bar{C_1}=\bar{C_2}=0$ imply that the
functional form of $C_1$ and $C_2$ must be such that they are
identically equal to zero at this critical point, otherwise this
solution does not exist.
 \be\label{47}
 {\rm solution \ \ 2}:\{ {\bar{x}}^2\geq 1,\
 \bar{y}=\bar{z}=\bar{u}=0,\
 \bar{v}=\pm\sqrt{{\bar{x}}^2-1}, \ \bar{C_1}=\bar{C_2}=0\}.
 \ee
In this case, the equations $\bar{C_1}=\bar{C_2}=0$ can generally
determine the allowed value of $\bar{x}$. If $\bar{C_1}$ and
$\bar{C_2}$ are identically equal to zero at
$\bar{y}=\bar{z}=\bar{u}=0$, as they are in $C=0$ case, then
$\bar{x}$ can choose any arbitrary value.
 \be\label{48}
 {\rm solution \ \ 3}:\{ {\bar{x}}=\bar{y}=0, \ \bar{z}=1, \ \bar{u},\
 \bar{v}=0, \ \bar{C_1}=\bar{C_2}=0\}.
 \ee
The value of $\bar{u}$ is generally determined by solving
 $\bar{C_1}=\bar{C_2}=0$.
 \be\label{49}
 {\rm solution \ \ 4}:\{ {\bar{x}}=0,\ \bar{y}=1, \ \bar{z}=0, \ \bar{u},\
 \bar{v}=0, \ \bar{C_1}=-\sqrt{\frac{3}{2}}\bar{f},\
 \bar{C_2}=0\},
 \ee
where the last two equations can generally determine $\bar{u}$.
 \be\label{50}
 {\rm solution \ \ 5}:\{ {\bar{y}}=\bar{u}=\bar{v}=0,
 \ \bar{C_1}=3\bar{x}({\bar{x}}^2+\frac{\gamma}{2}{\bar{z}}^2-1),\
 \bar{C_2}=-3\bar{z}({\bar{x}}^2+\frac{\gamma}{2}{\bar{z}}^2-\frac{\gamma}{2}),\
 {\bar{x}}^2+{\bar{z}}^2=1\}.
 \ee
 $\bar{x}$ and $\bar{z}$ are found by solving the above equations.
The last solution is:
 \bea\label{51}
 {\rm solution \ \ 6}:&\{ & {\bar{y}}=\sqrt{1-{\bar{x}}^2-{\bar{z}}^2},\ \bar{u}=\bar{v}=0, \
 \bar{C_1}=3\bar{x}({\bar{x}}^2+\frac{\gamma}{2}{\bar{z}}^2-1)-\sqrt{\frac{3}{2}}{\bar{y}}^2\bar{f},
 \cr && \bar{C_2}=-3\bar{z}({\bar{x}}^2+\frac{\gamma}{2}{\bar{z}}^2-\frac{\gamma}{2}),\
 3({\bar{x}}^2+\frac{\gamma}{2}{\bar{z}}^2)+\sqrt{\frac{3}{2}}{\bar{x}}\bar{f}=0
 \}.
 \eea

At $C=0$, solutions 1, 2, 3, 4, and 6 reduce to I.1, I.2, IV, III,
and II of Table \ref{tab1}, respectively. The solutions 1, 2, and
4 are hessence-dominated, solution 3 is
background-matter-dominated and 5 and 6 are generally scaling
solutions. Among these solutions, there are only two solutions 3
and 4 in which $\bar{u}$ can principally be different from zero,
which have not been appeared in examples discussed in \cite{17}.

In \cite{17}, besides the non-interacting $C=0$ case, three
following interactions have been considered:
 \bea\label{52}
 && C_1^{{\rm (II)}}=\sqrt{\frac{3}{2}}\alpha z^2,\cr
 && C_1^{{\rm (III)}}=\frac{3}{2}\frac{\beta}{x}, \cr
 && C_1^{{\rm (IV)}}=\frac{3}{2}\eta\frac{z^2}{x},
 \eea
in which $\alpha$, $\beta$ and $\eta$ are some constants. In both
of the solutions 3 and 4, $\bar{x}$ is zero, so $ C_1^{{\rm
(III)}}$ diverges and $ C_1^{{\rm (IV)}}$ is not generally
well-defined and therefore these critical points do not exist in
these cases. So we only consider the $C_1^{{\rm (II)}}$ case. For
$V=V_1$ potential, with $f_1=-\l$, solution 3 does not exist since
$\bar{C_1}=0$ leads to $\a =0$ which is not acceptable, and
solution 4 does also not exist as
$\bar{C_1}=-\sqrt{\frac{3}{2}}\bar{f}$ results in $\l =0$ which
again is not acceptable. For $V=V_2$ potential, with
$f_2=nu/\sqrt{6}$, solution 3 leads to $\alpha =0$ which is not
acceptable, and solution 4 results in $\bar{u}=0$, which does not
lead to a $\bar{u}\neq 0$ solution.

It may be useful to reproduce all the critical points of at least
one of the cases studied in \cite{17} in more detail. We consider,
as an example, $V=V_1$ and $C=C_1^{{\rm (II)}}$. So
 \be\label{53}
 f=-\l, \ \ \ C_1=\sqrt{\frac{3}{2}}\alpha z^2,\ \ \
 C_2=\sqrt{\frac{3}{2}}\alpha xz.
 \ee
The solutions 1-6 result in:
 \bea
 {\rm solution \ \  1}&:&\{ \bar{x}=\frac{\l}{\sqrt{6}},\
 \bar{y}=\sqrt{1-\frac{\l^2}{6}},\ \bar{z}=\bar{u}=
 \bar{v}=0\} , \label{54}\\
  {\rm solution \ \ 2}&:&\{ {\bar{x}}^2\geq 1,\
 \bar{y}=\bar{z}=\bar{u}=0,\
 \bar{v}=\pm\sqrt{{\bar{x}}^2-1}\} , \label{55}\\
 {\rm solution \ \ 3}&:& \ \ \bar{C_1}=0\rightarrow \alpha =0 \ \
 ({\rm is \ \ not \ \ acceptable}), \label{56}\\
 {\rm solution \ \  4}&:& \ \ \bar{C_1}=-\sqrt{\frac{3}{2}}\bar{f}  \rightarrow \l =0 \ \
 ({\rm is \ \ not \ \ acceptable}), \label{57}\\
 {\rm solution \ \ 5}&:&\{ \bar{x}=\sqrt{\frac{2}{3}}\frac{\alpha}{\gamma
 -2},\bar{y}=0, \bar{z}=\sqrt{1-\frac{2\alpha^2}{3(\gamma -2)^2}},\bar{u}=
 \bar{v}=0\} , \label{58}\\
 {\rm solution \ \ 6}&:&\{ \bar{x}=\sqrt{\frac{3}{2}}\frac{\gamma}{\l +\alpha }
 ,\bar{y}=\sqrt{\frac{2\alpha^2-3(\gamma -2)\gamma +2\alpha\l}{2(\alpha +\l )^2}},
  \bar{z}=\sqrt{\frac{\l (\l +\alpha )-3\gamma}{(\l +\alpha)^2}},\bar{u}=
 \bar{v}=0\} ,\cr && \label{59}
 \eea
which coincide with those in Table II of \cite{17}. Note that in
that table, four of the solutions ( 2p, 2m, 4 and 5) are not
independent solutions and are special cases of the first solution.

The stability studies of these critical points depends on the
precise value of the function $C$. But it may be interesting to
study the conditions under which the derivative of $f$ becomes
important in the stability properties of the critical points.
Consider the most general case $C_1=C_1(x,z,u)$. It can be shown
that the coefficient of $\bar{f'}$ term in equation
$\det(M-\hat{1}\l)=0$, where $\hat{1}$ stands for $4\times 4$ unit
matrix, is
  \be\label{60}
  \frac{{\bar{y}}^2{\bar{u}}^2}{{\bar{z}}^2}h(\bar{x},\bar{y},\bar{z},\bar{u}),
  \ee
which shows that the derivative of $f$ survives only if
  \be\label{61}
  \bar{y}\neq 0 \ \ {\rm and} \ \ \bar{u}\neq 0.
  \ee
Therefore the solution 4 is the only one in which we expect
$\bar{f'}$ may be important. Focusing on this solution, the
eigenvalue problem is still very complicated in general case and
it is better to consider the more restricted cases. As an example,
we consider the potentials and $C$'s in which
$\bar{f}=\overline{(C_1/z)}=\overline{(\partial C_1/\partial
x)}=0$. Under these conditions, the equation $\det(M-\hat{1}\l)=0$
leads to
  \be\label{62}
  \l^4+a_1\l^3+a_2\l^2+a_3\l +a_4=0,
  \ee
where
  \bea\label{63}
  && a_1=9+\frac{3}{2}\gamma, \cr
  && a_2=18+\frac{27}{2}\gamma -{\bar{u}}^2(\bar{C}_{1,u}+
  \sqrt{\frac{3}{2}}\bar{f'}), \cr
  &&  a_3=-3{\bar{u}}^2(\bar{C}_{1,u}+\sqrt{\frac{3}{2}}
 \bar{f'})(2+\frac{\gamma}{2})+27\gamma, \cr
  &&  a_4=-9\gamma{\bar{u}}^2(\bar{C}_{1,u}+\sqrt{\frac{3}{2}}
 \bar{f'}),
  \eea
 where $\bar{C}_{1,u}=\overline{(\partial C_1/\partial
u)}$. Now it is well known that $\sum_i\l_i=-a_1$, $\sum_{i\neq
j}\l_i\l_j=a_2$,  $\sum_{i\neq j\neq k}\l_i\l_j\l_k=-a_3$, and
$\l_1\l_2\l_3\l_4=a_4$, where $\l_1\cdots\l_4$ are the roots of
eq.(\ref{62}). So all the roots are non-positive only if $a_i\geq
0$ $(i=1,\cdots ,4)$, which provided if
  \be\label{64}
   \bar{f'}\leq -\sqrt{\frac{2}{3}}\bar{C}_{1,u}.
   \ee
In $C=0$, this condition of stability reduces to eq.(\ref{43}).

\section{Conclusion}
In this paper, we study the attractor solutions of the general
hessence model by studying the 4-dimensional phase space of the
theory. The hessence model is a non-canonical complex scalar
theory which can be a candidate of dark energy with some
interesting properties, among them is the possibility of crossing
the $w=-1$ line. Comparing the Lagrangian of hessence model with
quintom model shows that the $\theta$-( or equally charge
$Q$-)term plays the role of the phantom field. In $Q=0$, the model
reduced to quintessence model with no $w=-1$ crossing.

We consider an arbitrary hessence potential $V(\phi)$ and almost
arbitrary hessence-background matter interaction term $C$, and
find several results. We show that in $C=0$, there is always at
least one stable attractor which depends on the value
$\bar{f}=f(\bar{u}=0)$: For ${\bar{f}}^2\leq 3\gamma$ the
hessence-dominated attractor I.1 and for ${\bar{f}}^2\geq 3\gamma$
the scaling attractor II of Table 1. In all the attractor
solutions we have $\bar{\phi}\rightarrow\infty$, except for the
attractor III where $\bar{\phi}$ is finite and also its stability
depends on the derivative $(df/du)_{\bar{u}}$. This kind of
attractor did not appear in the previously studied cases and can
be seen in potentials like $\sin(\kappa\phi)$. $\bar{v}$ is zero
in all the stable attractors of Table 1, but for arbitrary
$C$-term, this is not the case.

For general $C$, we show that there generally exist six classes of
attractor, which all of them can be stable in special cases ( for
example $V=V_1$ and $C_1=C_1^{(II)}$ of \cite{17}). Among these
solutions, solution 2 is the only one with property $\bar{v}\neq
0$. The solutions 3 and 4 have finite $\bar{\phi}$ value and the
the solution 4 is the only one which its stability depends on
$(df/du)_{\bar{u}}$, the features that can not be seen in the
previously studied potentials.

 {\bf Acknowledgement:}  We
would like to thank the research council of the University of
Tehran for partial financial support.

\end{document}